\documentclass{PoS}
\usepackage{graphicx}

\title{\boldmath Search for lepton-flavor-violating $\tau$
 decay and lepton-number-violation $B$ decay at Belle}

\ShortTitle{Search for $\tau$ LFV and $B$ LNV at Belle}

\author{\speaker{K.Hayasaka}\thanks{the Belle collaboration}\\
        Nagoya University\\
        E-mail: \email{hayasaka@hepl.phys.nagoya-u.ac.jp}}

\abstract{We report on the recent results
of searches for lepton-flavor-violating $\tau$ decays
as well as lepton-number-violating $B$ decays
with the world-highest data samples
accumulated at the Belle.}

\FullConference{35th International Conference of High Energy Physics - ICHEP2010,\\
		July 22-28, 2010\\
		Paris France}

\begin{document}
\section{Introduction}
An observation of the lepton-flavor-violating (LFV)
or the lepton-number-violating (LNV) processes
is a clear signature of the existence for physics 
beyond the standard model (BSM) since
they are forbidden in the standard model (SM).
Among various LFV processes,
some LFV decays of tau lepton are expected to be
enhanced by several kinds of BSM, especially, 
models with supersymmetry. One of the most frequently 
discussed is $\tau^-\rightarrow\mu^-\gamma$ decay.
 But, in some models, such as
the non-universal Higgs mass model,
the constrained minimal supersymmetric SM and so on,
among the $\tau$ LFV decays,
$\tau^-\rightarrow\mu^-\eta$ or $\tau^-\rightarrow\mu^-\rho^0$
will be enhanced.~\cite{Arganda:2008jj}
In this talk, we report the recent result
of a search for $\tau^-\rightarrow\ell^- M^0$ ($\ell=e,\mu$,
$M^0=\pi^0,\eta,\eta^{\prime},\rho^0,K^{\ast0},\bar{K}^{\ast0},\omega$
and $\phi$) with the world largest data sample accumulated at Belle.
On the other hand,
some LNV decays
are possible when the neutrino is the Majorana type spinor.
Here, we show the first result of search for $B^+\rightarrow D^- \ell^+
\ell^+$ at Belle.
\section{\boldmath $\tau\rightarrow\ell M^0$ ($\ell=e,\mu$, 
$M^0=\pi^0,\eta,\eta^{\prime},\rho^0,K^{\ast0},\bar{K}^{\ast0},\omega$
and $\phi$)}
\subsection{Analysis Method}
In the $\tau$ LFV analysis, in order to evaluate the number of
 signal events, two independent variables are defined,
that are
signal-reconstructed mass
and energy in the center-of-mass (CM) frame
from energies and momenta for the signal $\tau$ daughters. 
In the $\tau\rightarrow\mu\eta$ case,
they are defined as
$M_{\mu\eta}=\sqrt{E_{\mu\eta}^2-P_{\mu\eta}^2},\ 
 \Delta E=E_{\mu\eta}^{\rm CM}-E_{\rm beam}^{\rm CM}$,
where $E_{\mu\eta}$ ($P_{\mu\eta}$) is a sum of the energies 
(a magnitude of a vector sum of the momenta) for $\mu$ and $\eta$,
the superscript $\rm CM$ indicates that the variable is defined in the
CM frame and $E_{\rm beam}^{\rm CM}$ 
means the initial beam energy in the CM frame.
Principally, $M_{\mu\eta}$ and $\Delta {E}$ should be
$m_{\tau}$ ($\sim 1.78$ GeV/$c^2$) and 0 (GeV), respectively,
for signal events while $M_{\mu\eta}$ and $\Delta {E}$ will
smoothly vary without any special structure in the background (BG)
events. Due to a finite resolution, the signal events are distributed
 around $M_{\mu\eta}\sim m_{\tau}$ and $\Delta E\sim0$ (GeV).
Taking into account the resolution, we set the elliptic signal region.
Finally,
we evaluate the number of signal events in the signal region.
When the number of the observed events is consistent to
that of the expected BG events, the upper limit for the
number of the signal events is evaluated by Feldman-Cousins
method.~\cite{Feldman:1997qc}
To avoid any bias, we perform the blind analysis:
Before fixing the selection criteria and the evaluation 
for the systematic uncertainties, we cover the data 
events in the signal region.

\subsection{\boldmath $\tau\rightarrow\ell P^0$ ($\ell=e,\mu$,
  $P^0=\pi^0,\eta,\eta^{\prime}$)}
\begin{figure}[t]
\vspace*{-6mm}
\includegraphics[width=\textwidth]{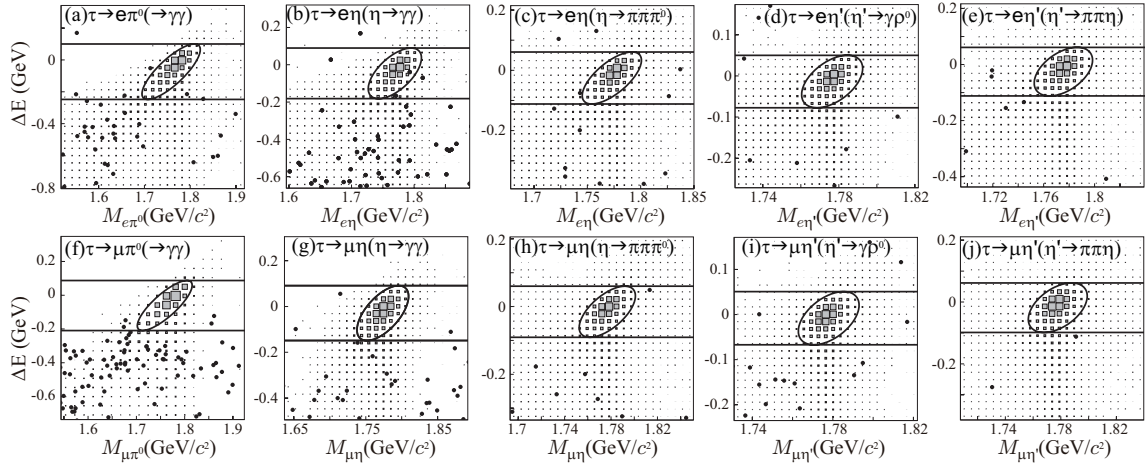}
\caption{\small Resulting 2D plots for 
$\tau\rightarrow e \pi^0$~(a),
$e \eta(\rightarrow\gamma\gamma)$~(b),
$e \eta(\rightarrow\pi\pi\pi^0)$~(c),
$e \eta^\prime(\rightarrow\gamma\rho^0(\rightarrow\pi\pi))$~(d),
$e \eta^\prime(\rightarrow\pi\pi\eta(\rightarrow\gamma\gamma))$~(e),
$\mu \pi^0$~(f),
$\mu \eta(\rightarrow\gamma\gamma)$~(g),
$\mu \eta(\rightarrow\pi\pi\pi^0)$~(h),
$\mu \eta^\prime(\rightarrow\gamma\rho^0(\rightarrow\pi\pi))$~(i)
and
$\mu \eta^\prime(\rightarrow\pi\pi\eta(\rightarrow\gamma\gamma))$~(j),
on the $M_{\ell P^0}$ -- $\Delta E$ plane. Here,
black dots (shaded boxes) express the data (signal MC),
the region bounded by two lines is defined
as a $3\sigma$ band for the BG estimation
and the elliptic region is the signal one which corresponds
to $3\sigma$
in each plot. One event is found in the signal region for
$\tau\rightarrow e \eta$
while no events appear in any other modes.\vspace*{-3mm}}
\label{fig:lp0result}
\end{figure}

We perform a new search for the $\tau$ decay into a lepton ($e$ or
$\mu$) and a neutral pseudoscalar ($\pi^0$, $\eta$ or $\eta^{\prime}$)
with a 901 fb$^{-1}$ data sample.
A neutral pion is reconstructed from 2 photons
while an $\eta$ ($\eta^{\prime}$)
is reconstructed from $\gamma\gamma$ ($\rho^0\gamma$)
as well as $\pi^+\pi^-\pi^0$ ($\pi^+\pi^-\eta$)
to increase the detection efficiency.
When the neutral pseudoscalars are reconstructed,
their four-momentum is evaluated by a mass-constrained fit
to obtain a better resolution for the signal region.
Because we have modified the selection criteria applied
to the previous analysis, we obtain an about $1.5$ times
better detection efficiency while the similar background 
level is kept. As a result, we observe one event
in the $\tau\rightarrow e\eta (\rightarrow\gamma\gamma)$ mode
while no events are found in other modes as shown 
in Fig.~\ref{fig:lp0result}. Since these results
are consistent with the background estimation,
we set upper limits 
on the following branching fractions:
${\cal{B}}(\tau^-\rightarrow e^-\pi^0) < 2.7\times 10^{-8}$,
${\cal{B}}(\tau^-\rightarrow \mu^-\pi^0) < 2.2\times 10^{-8}$,    
${\cal{B}}(\tau^-\rightarrow e^-\eta) < 4.4\times 10^{-8}$, 
${\cal{B}}(\tau^-\rightarrow \mu^-\eta) < 2.3\times 10^{-8}$,
${\cal{B}}(\tau^-\rightarrow e^-\eta') < 3.6\times 10^{-8}$
and 
${\cal{B}}(\tau^-\rightarrow \mu^-\eta') < 3.8\times 10^{-8}$,  
at the 90\% confidence level.
\begin{figure}[b]
\vspace*{-3mm}
\includegraphics[width=\textwidth]{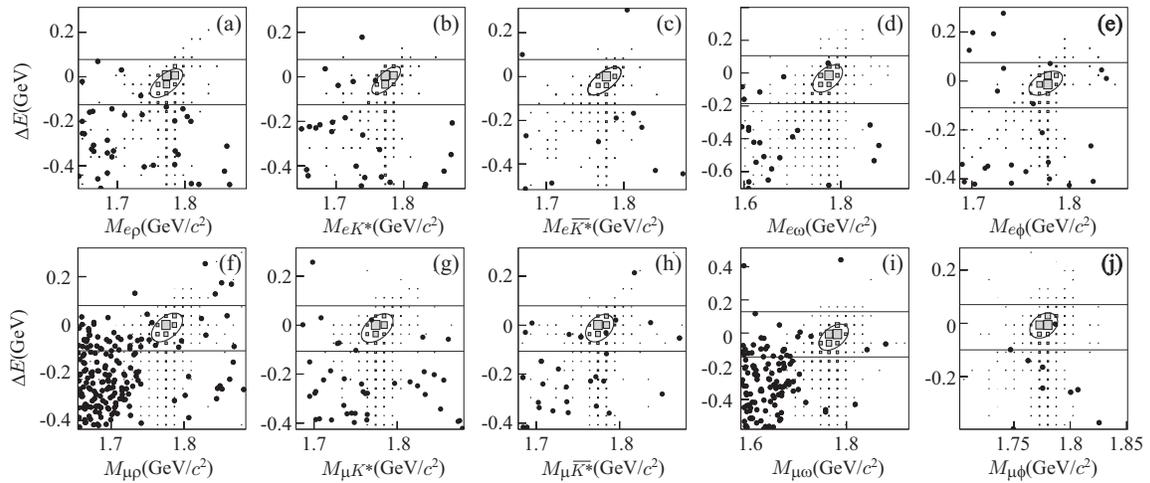}
\caption{\small Resulting 2D plots for 
$\tau\rightarrow e \rho^0$~(a),
$e K^{\ast0}$~(b),
$e \bar{K}^{\ast0}$~(c),
$e\omega$~(d),
$e\phi$~(e),
$\mu \rho^0$~(f),
$\mu K^{\ast0}$~(g),
$\mu \bar{K}^{\ast0}$~(h),
$\mu\omega$~(i) and
$\mu\phi$~(j)
on the $M_{\ell V^0}$ -- $\Delta E$ plane. Here,
black dots (shaded boxes) express the data (signal MC),
the region bounded by two lines is defined
as a $5\sigma$ band for the BG estimation
and the elliptic region is the signal one which corresponds
to $3\sigma$
in each plot. One event is found in the signal region for
$\tau\rightarrow\mu K^{\ast 0}$, $\mu \bar{K}^{\ast0}$ and $\mu\phi$
while no events appear in any other modes.\vspace*{-3mm}}
\label{fig:lv0result}
\end{figure}
\subsection{\boldmath $\tau\rightarrow\ell V^0$
($\ell=e,\mu$, 
$V^0=\rho^0,K^{\ast0},\bar{K}^{\ast0},\omega,\phi$)}
Similarly to $\tau^-\rightarrow\ell^- P^0$, we update our results
of the search for $\tau^-\rightarrow\ell^- V^0$, 
where 
$\ell=e,\mu$, $V^0=\rho^0,K^{\ast0},\bar{K}^{\ast0},\omega,\phi$
with an 854 fb${}^{-1}$ data sample. By performing a detailed
background study, we obtain a 1.2 times better efficiency
in average with keeping similar level backgrounds.
Finally, one event is found in the signal region for
$\tau^-\rightarrow\mu^- K^{\ast 0}$, $\mu^- \bar{K}^{\ast0}$ 
and $\mu^-\phi$
while no events appear in any other modes.
They are consistent with the expected number of the backgrounds.
Consequently, we set the 90\% confidence level upper limits
on the branching fractions:
${\cal B}(\tau^-\rightarrow e^-\rho^0)<1.8\times10^{-8}$,
${\cal B}(\tau^-\rightarrow e^- K^{\ast0})<3.2\times10^{-8}$,
${\cal B}(\tau^-\rightarrow e^- \bar{K}^{\ast0})<3.4\times10^{-8}$,
${\cal B}(\tau^-\rightarrow e^- \omega)<4.8\times10^{-8}$,
${\cal B}(\tau^-\rightarrow e^- \phi)<3.1\times10^{-8}$,
${\cal B}(\tau^-\rightarrow \mu^-\rho^0)<1.2\times10^{-8}$,
${\cal B}(\tau^-\rightarrow \mu^- K^{\ast0})<7.2\times10^{-8}$,
${\cal B}(\tau^-\rightarrow \mu^- \bar{K}^{\ast0})<7.0\times10^{-8}$,
${\cal B}(\tau^-\rightarrow \mu^- \omega)<4.7\times10^{-8}$,
${\cal B}(\tau^-\rightarrow \mu^- \phi)<8.4\times10^{-8}$.

\section{\boldmath $B^+\rightarrow\ell^+\ell^{\prime+}D^-$ 
($\ell,\ell^{\prime}=e,\mu$)}
\noindent
\begin{minipage}[b]{0.58\textwidth}
\ \ 
If the neutrino is a Majorana-type spinor,
a $B^+\rightarrow\ell^+\ell^{\prime+}h^-$ decay is possible,
where $\ell,\ell^{\prime}=e,\mu$, $h=\pi,K,\rho,K^{\ast},D,\cdots$
~\cite{Cvetic:2010rw}.
Due to the size of CKM matrix, 
$B^+\rightarrow\ell^+\ell^{\prime+}D^-$
is expected to have the largest branching fraction
among them. But this mode has never been measured yet.
Using a $7.7\times10^8$ $B^+B^-$ data sample,
we perform a first search.
Similarly to the $\tau$ LFV case,
we define $M_{\rm bc}$ and $\Delta E$,
where they are reconstructed from $\ell^+$,
$\ell^{\prime+}$ and $D^-$, but the total
energy is set to the initial beam energy.
In order to evaluate the number of signal events,
the signal box is defined as $5.27$ (GeV/$c^2$)
$<M_{\rm bc}$
\end{minipage}
\hspace*{0.02\textwidth}
\begin{minipage}[b]{0.4\textwidth}
\begin{center}
\addtocounter{figure}{1}
\includegraphics[width=\textwidth]{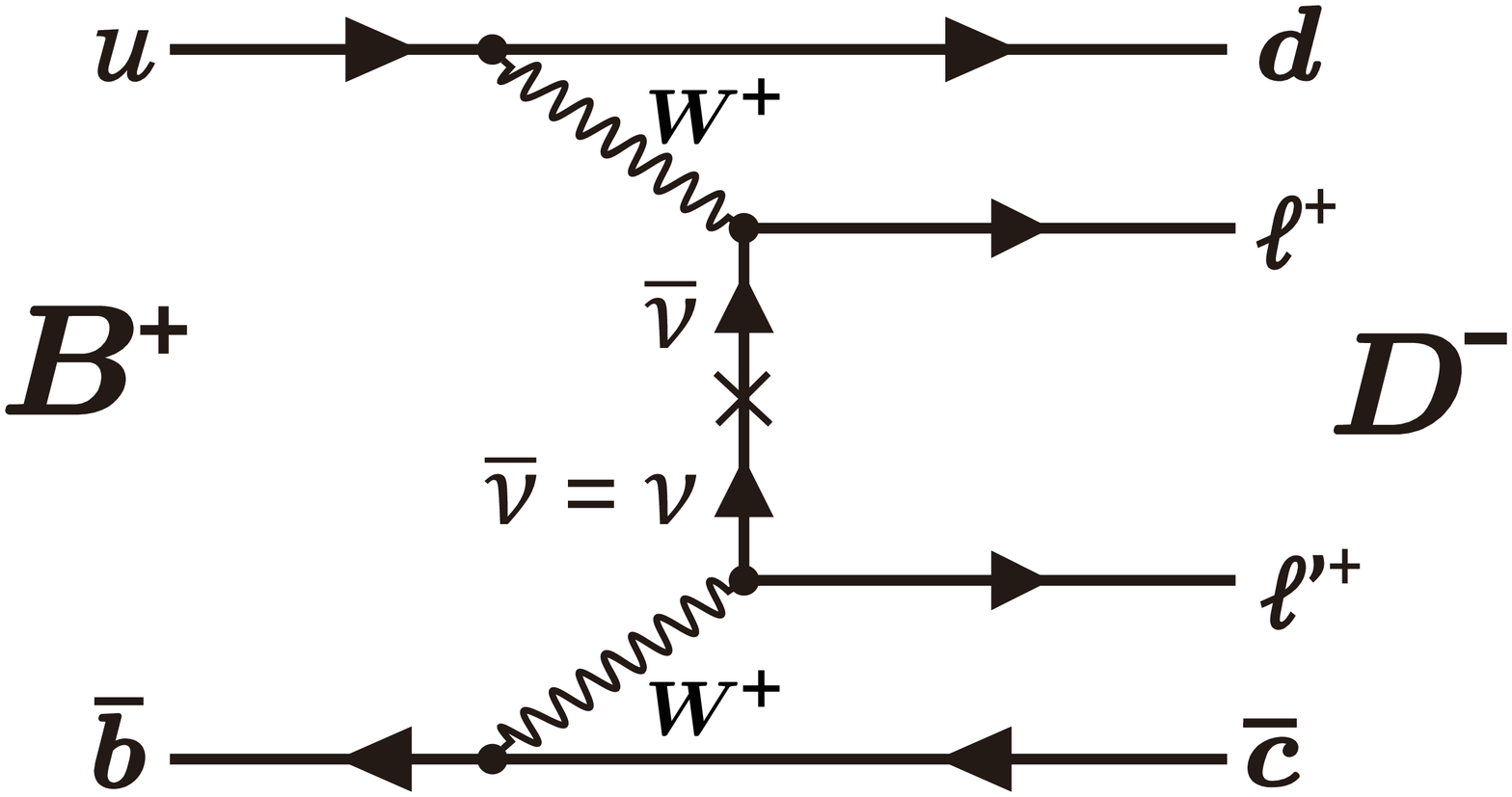}
\vspace*{-10mm}\\
\end{center}
{\small {\bf Figure \thefigure:} Decay process for $B^+\rightarrow\ell^+\ell^{\prime+}D^-$.
When a neutrino is a Majorana spinor, the neutrinos can connect
as an inner line in the process.}
\end{minipage}
$<$ 5.29 (GeV/$c^2$)
and $-0.035$ (GeV) $<\Delta E<$ $0.035$ (GeV) for 
$B^+\rightarrow \mu^+\mu^+D^-$, or
$-0.055$ (GeV) $<\Delta E<$ 
$0.035$ (GeV) for 
others, because the distribution for the 
electron energy has a small tail on the lower side.
After the selection, we found no events in the signal box
for all modes.
Finally, we set the 90\% confidence level upper limits
on the branching fractions:
${\cal B}(B^+\rightarrow e^+ e^+D^-)<2.7\times10^{-6}$,
${\cal B}(B^+\rightarrow e^+ \mu^+D^-)<1.9\times10^{-6}$,
${\cal B}(B^+\rightarrow \mu^+ \mu^+D^-)<1.1\times10^{-6}$.

\begin{figure}[h]
\includegraphics[width=\textwidth]{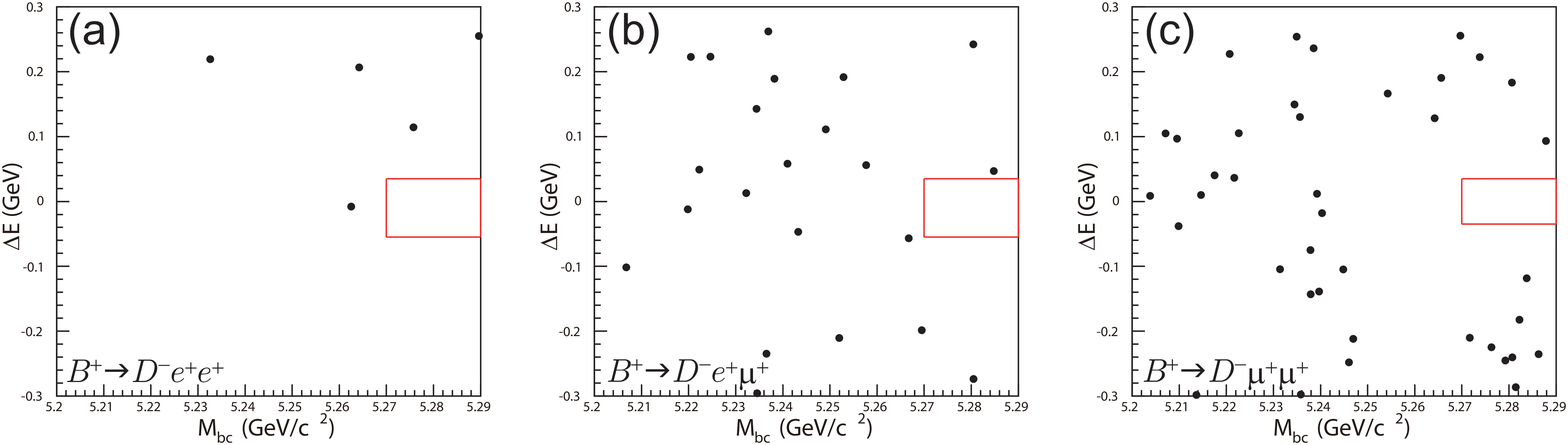}
\caption{Resulting 2D plots for 
$B^+\rightarrow e^+e^+D^-$ (a),
$B^+\rightarrow e^+\mu^+D^-$ (b) and 
$B^+\rightarrow \mu^+\mu^+D^-$ (c) 
on the $M_{\rm bc}$ -- $\Delta E$ plane. Here,
black dots express the data and the box region is the signal one
in each plot. No events are found in each mode.\vspace*{-3mm}}
\label{fig:lnvresult}
\end{figure}


\begin{thebibliography}{99}
\bibitem{Arganda:2008jj}
  E.~Arganda, M.~J.~Herrero and J.~Portoles,
  JHEP {\bf 0806}, 079 (2008).
\bibitem{Feldman:1997qc}
  G.~J.~Feldman and R.~D.~Cousins,
  Phys.\ Rev.\  D {\bf 57}, 3873 (1998).
\bibitem{Cvetic:2010rw}
  G.~Cvetic, C.~Dib, S.~K.~Kang and C.~S.~Kim,
  Phys.\ Rev.\  D {\bf 82}, 053010 (2010).
\end{thebibliography}
\end{document}